\documentstyle[prd,aps,epsfig]{revtex}

\newcommand{\beq}{\begin{equation}}
\newcommand{\eeq}{\end{equation}}
\newcommand{\baq}{\begin{eqnarray}}
\newcommand{\eaq}{\end{eqnarray}}

\begin{document}
\title{ Parametrization of a nonlocal, chiral quark model \\
in the instantaneous three-flavor case. \\
Basic formulas and tables}
\author{Hovik Grigorian \thanks{%
Permanent address: Department of Physics, Yerevan State
University, 375047 Yerevan, Armenia}}
\address{Instit\"ut f\"ur Physik, Universit\"at Rostock, D-18051 Rostock,
Germany}
\date{\today}
\maketitle

\begin{abstract}
In these notes we describe the basic formulation of the
parametrization scheme for the instantaneous nonlocal chiral quark
model in the three flavor case. We choose to discuss the Gaussian,
Lorentzian-type, Woods-Saxon and sharp cutoff (NJL) functional
forms of the momentum dependence for the formfactor of the
separable interaction. The four parameters: light and strange
quark mass, coupling strength ($G_S$) and range of the interaction
($\Lambda$) have been fixed by the same phenomenological inputs:
pion and kaon mass, pion decay constant and light quark mass in
vacuum. The Wood-Saxon and Lorentzian-type formfactor is suitable
for an interpolation between sharp cutoff and soft momentum
dependence. Results are tabulated for applications in models of
hadron structure and quark matter at finite temperatures and
chemical potentials where separable models have proven successful.
\end{abstract}


\preprint{\parbox{5cm}{MPG--VT--UR 271/06}}

\section{Introduction}

The phase diagram of  Quantum Chromodynamics (QCD) is investigated
in large-scale lattice gauge theory simulations
\cite{Karsch:2006sm} and heavy-ion collision experiments at
CERN-SPS and RHIC Brookhaven \cite{Muller:2006ee}, where the
approximately baryon-free region at finite temperatures is
accessible and consensus about the critical temperature for the
occurence of a strongly correlated quark-gluon plasma phase (sQGP)
is developing. The region of low temperatures and high baryon
densities, however, which is interesting for the astrophysics of
compact stars, is not accessible to lattice QCD studies yet and
heavy-ion collision experiments such as the CBM experiment at FAIR
Darmstadt are still in preparation \cite{Senger:2006}. The most
stringent of the presently available constraints on the EoS of
superdense hadronic matter from compact stars and heavy-ion
collisions have recently been discussed in
Ref.~\cite{Klahn:2006ir} and may form the basis for future
systematic investigations of the compatibility of dense quark
matter models with those phenomenological constraints. Therefore,
the question arises for appropriate models describing the
nonperturbative properties of strongly interacting matter such as
dynamical chiral symmetry breaking and hadronic bound state
formation in the vacuum and at finite temperatures and densities.

The Nambu--Jona-Lasinio (NJL) model has proven very useful for
providing results to this question within a simple, but
microscopic formulation, mostly on the mean-field level, see
\cite{Buballa:2003qv}. The state of the art phase diagrams of
neutral quark matter for compact star applications have recently
been obtained in
\cite{Ruster:2005jc,Blaschke:2005uj,Aguilera:2004ag,Abuki:2005ms}
where also references to other approaches can be found. One of the
shortcomings of the NJL model is the absence of confinement, the
other is its nonrenormalizability. It is customary to speak of the
NJL model in its form with a cut-off regularization, where
physical observables can be defined and calculated. The cutoff in
momentum space, however, defines a range of the interaction and
makes the NJL model nonlocal. It has been suggested that the
cutoff-regularized NJL model can be considered as a limiting case
of a more general formulation of nonlocal chiral quark models
using separable interactions \cite{Schmidt:1994di}. In this form
one can even make contact with the Dyson-Schwinger equation
approach to QCD by defining a separable representation of the
effective gluon propagator \cite{Blaschke:2000gd}, or to the
instanton liquid model, see \cite{GomezDumm:2005hy}.

We have made extensive use of the parametrization given in
\cite{Schmidt:1994di} for studies of quark matter phases in
compact stars \cite{Blaschke:2003yn,Grigorian:2003vi} where the
role of the smoothness of the momentum dependence for the
quark-hadron phase transition and compact star structure has been
explored. These investigations have been also used in simulations
of hybrid star cooling \cite{Grigorian:2004jq,Popov:2005xa}, which
can be selective for the choice of the quation of state (EoS) of
quark matter by comparing to observational data feor surface
temperature and age of compact stars. As a result of these
studies, color superconducting phases with small gaps of the order
of 10 keV - 1 MeV appear to be favorable for the cooling
phenomenology. A prominent candidate, the color-spin-locking (CSL)
phase, has been investigated more in detail within the NJL model
with satisfactory results  \cite{Aguilera:2005tg}. However, its
generalization to formfactors with a smooth momentum dependence
revealed a severe sensitivity resulting in variations of the CSL
gaps over four orders of magnitude \cite{Aguilera:2005uf}.

Unfortunately, with the NJL parametrization given in
\cite{Schmidt:1994di} it was not possible to reproduce results
with NJL parametrizations given in \cite{Buballa:2003qv} and used,
e.g., in  Refs.
\cite{Ruster:2005jc,Blaschke:2005uj,Aguilera:2005tg}. Therefore,
in the present work a new parametrization of the model presented
in Ref. \cite{Schmidt:1994di} is performed with a special emphasis
on reproducing NJL parametrizations given in \cite{Buballa:2003qv}
in the limiting case of a sharp cutoff formfactor. We also take
into account the strangeness degree of freedom and consider
Lorentzian-type formfactor models where the form of the momentum
dependence for the quark-quark interaction can be varied
parametrically thus being most suitable for a quantitative
analysis the phase diagram and high-density EoS under the above
mentioned constraints from compact star and heavy-ion collision
phenomenology.

\section{Basic formulation}

We consider a nonlocal chiral quark model with separable
quark-antiquark interaction in the color singlet
scalar/pseudoscalar isovector channel \cite{Gocke:2001ri} where
the formfactors are given in the  instantaneous approximation, in
the same way as it was suggested in \cite{Schmidt:1994di}.

The Lagrangian density of the quark model is given by
($i,j=u,d,s$)
\begin{equation}
{\cal L}= \bar{q}_{i}(i \gamma_\mu \partial^\mu - m_{i,0})q_{i } +
G_{S}\sum_{a=0}^{8}\left[ (\bar{q_i}\tilde{g}(x)~\lambda
_{ij}^{a}q_j)^{2}+(\bar{q_i}(i~\tilde{g}(x)\gamma _{5})\lambda
_{ij}^{a}q_j)^{2}\right],
\end{equation}
where indices occuring twice are to be summed over and the
formfactor $\tilde{g}(x)$ for the  nonlocal current-current
coupling has been introduced. Here $m_0=m_{u,0}=m_{d,0}$ and
$m_{s,0}$ are the current quark masses of the light and strange
flavors, respectively, $\lambda _{ij}^a$ are the Gell-Mann
matrices of the $SU(3)$ flavor group and $\gamma _{\mu}$, $\gamma
_5$ are Dirac matrices.

The nonlocality of the current-current interaction in the
quark-antiquark ($q \bar{q}$) channel is implemented in the
separable approximation via the same formfactor functions for all
colors and flavors. In our calculations we use the Gaussian (G),
Lorentzian (L), Woods-Saxon (WS) and cutoff (NJL) formfactors in
momentum space defined as (see Ref. \cite{Schmidt:1994di})
\begin{eqnarray}
g_{{\rm G}}(p) &=&\exp (-p^{2}/\Lambda _{{\rm G}}^{2})~,
  \nonumber \\
g_{{\rm L}}(p) &=&[1+(p/\Lambda _{{\rm L}})^{2\alpha }]^{-1},
\nonumber \\
g_{{\rm WS}}(p)&=& [1+\exp(- \alpha)]/\{1+\exp[\alpha~
(p^2/\Lambda_{{\rm WS}}^2-1]\},\nonumber \\
 g_{{\rm NJL}}(p) &=&\theta (1-p/\Lambda
_{{\rm NJL}})~.\nonumber
\end{eqnarray}
The formfactors can be introduced in a manifestly covariant way
(see \cite{Gocke:2001ri}), but besides technical complications at
finite $T$ and $\mu$, where Matsubara summations have to be
performed numerically, it is not a priori obvious that such a
formulation shall be superior to an instantaneous approximation
(3D) which could be justified as a separable representation of a
Coulomb-gauge potential model \cite{Blaschke:1994px}.

Typically, three-flavor NJL type models use a 't Hooft determinant
interaction that induces a U$_{A}$(1) symmetry breaking in the
pseudoscalar isoscalar meson sector, which can be adjusted such
that the $\eta $-$\eta^{\prime }$ mass difference is described. In
the present approach this term is neglected using the motivation
given in \cite{Blaschke:2005uj}, so that the flavor sectors
decouple in the mean-field approximation.

The dynamical quark mass functions are then given by
$M_i(p)=m_{i,0}+\phi _i~g(p)$, where the chiral gaps fulfill the
gap equations
\begin{eqnarray}
\phi_{i} &=& 4G_{S}\frac{N_{c}}{\pi ^{2}}\int dp
p^{2}g(p)\frac{M_{i}(p)}{E_{i}(p)}~, \label{disprel}
\end{eqnarray}
corresponding to minima of the thermodynamic potential with
respect to variations of the order parameters $\phi_{i}$, the
quark dispersion relations are $E_{i}(p)=\sqrt{p^{2}+M_{i}^2(p)}$.

The basic set of equations should be chosen to fix the parameters
included in the model, which are the current masses, coupling
constant and cutoff parameter ($m_{0},m_{s,0},G_{S}$ and
$\Lambda$).

In order to do that we use the properties of bound states of
quarks in the vacuum given by the pion decay constant $f_{\pi
}=92.4$~MeV, the masses of the pion $M_{\pi }=135$ MeV and the
kaon $M_{K}=494$ MeV and either the constituent quark mass
$M(p=0)=m_0+\phi_u$ or the chiral condensate of light quarks,
defined as
\begin{equation}
\langle u \bar{u}\rangle _0 =-\frac{N_{c}}{\pi ^{2}}\int dp
p^{2}\frac{M_{u}(p)-m_{0}}{E_{u}(p)}~, \label{condensate}
\end{equation}
with a phenomenological value from QCD sum rules \cite{Dosch} of
$190$ MeV $\le -\langle u \bar{u}\rangle^{1/3}_0\le 260$ MeV. The
chiral condensate generally is not properly defined in the case of
nonlocal interactions. The subtraction of the $m_{0}$ term has
been included to make the integral convergent.

The pion ~decay ~constant can be expressed in the form
\begin{equation}
\label{fpi} f_{\pi }=\frac{3~g_{\pi q\bar{q}}}{2\pi ^{2}} \int dp
p^{2}g(p)\frac{M_{u}(p)}{E_{u}(p)(E_{u}(p)^{2}-M_{\pi }^{2}/4)},
\end{equation}%
where the pion wave function renormalization factor $g_{\pi
q\bar{q}}$  is
\[
g_{\pi q\bar{q}}^{-2}= \frac{3}{2\pi ^{2}}\int dp p^{2}g^{2}(p)
\frac{E_{u}(p)}{(E_{u}(p)^{2}-M_{\pi }^{2}/4)^{2}}.
\]
The masses of pion and kaon are obtained from a direct
generalization of the well-known NJL model
\cite{Rehberg:1995kh,Costa:2005cz} by introducing formfactors with
the momentum space integration and replacing constituent quark
masses by the momentum dependent mass functions $M(p)$
\begin{eqnarray}
\label{Mpi} M_{\pi } &=&\left[ \left(
\frac{1}{2G_{S}}-2I_{u}^{(1)}\right) /I_{uu}^{(2)}
\right] ^{1/2}, \\
M_{K} &=&\left[ \left(
\frac{1}{2G_{S}}-(I_{u}^{(1)}+I_{s}^{(1)})\right)
/I_{us}^{(2)}\right] ^{1/2},
\end{eqnarray}
In these mass formulae, the following abbreviations for integrals
have been used
\begin{eqnarray}
I_{i}^{(1)} &=&\frac{3}{\pi ^{2}}\int dp
p^{2}g^{2}(p)\frac{1}{E_{i}(p)},
\nonumber \\
I_{uu}^{(2)} &=&\frac{3}{2\pi ^{2}} \int dp
p^{2}\frac{g^{2}(p)}{E_{u}(p)(E_{u}(p)^{2}- M_{\pi }^{2}/4)},
\nonumber \\
I_{us}^{(2)} &=&\frac{3}{\pi ^{2}}\int dp p^{2} g^{2}(p)
\frac{E_{u}(p)+E_{s}(p)}{E_{u}(p)E_{s}(p)[(E_{u}(p)+E_{s}(p))^{2}-M_{K}^{2}]}.
\end{eqnarray}
We can use these notations to give an estimate of the validity of
low-energy theorems for this nonlocal generalization of the NJL
model. To this end we rewrite Eq. (\ref{fpi}) for $f_\pi$ and
$g_{\pi q\bar{q}}$ as

\begin{eqnarray}
\label{fpi-2} f_{\pi }&=&g_{\pi q\bar{q}}\left(\phi_u~
I_{uu}^{(2)}
+ m_0 \langle g^{-1}(p) \rangle^{(2)}\right)\\
g_{\pi q\bar{q}}^{-2}&\approx& I_{uu}^{(2)} + M_{\pi }^{2}/4 \cdot
\langle E^{-2}_{u}(p) \rangle^{(2)}~, \label{gpi-2}
\end{eqnarray}
where the mean values of a distribution $F(p)$ are defined using
the integral $I^{(2)}_{uu}$ as an operator: $\langle
F(p)\rangle^{(2)}=I^{(2)}_{uu}[F(p)]$. To leading order in an
expansion at the chiral limit ($m_0\to 0$, $M_\pi\to 0$) one
obtains the Goldberger-Treiman relation
\begin{equation}
\label{GT} f_{\pi }g_{\pi q\bar{q}}=\phi_u~.
\end{equation}
Rewriting the gap equation (\ref{disprel}) for the light flavor as
\begin{equation}
\label{gap} \phi_u[1-4G_S~I_{u}^{(1)}]=- m_0 \langle u\bar{u}
\rangle_0 \cdot 4 G_S
\end{equation}
and the pion mass formula (\ref{Mpi}) as
\begin{equation}
\label{Mpi2} M_{\pi }^2 =\frac{1}{2G_{S}}\left(1 -4
G_S~I_{u}^{(1)}\right) /I_{uu}^{(2)}~,
\end{equation}
we obtain by combining (\ref{Mpi2}) with (\ref{gap}), (\ref{GT})
and (\ref{fpi-2}) in leading order the Gell-Mann--Oakes--Renner
relation (GMOR)
\begin{equation}
\label{GMOR} M_{\pi }^{2}f_{\pi }^{2}=- 2 m_{0}\langle u\bar{u}
\rangle_0~.
\end{equation}
As an indicator of the validity of this low-energy theorem we will
show the GMOR value for the light current quark mass
\begin{equation}
m_{0}^{GMOR}=-\frac{M_{\pi }^{2}f_{\pi }^{2}}{2\langle u
\bar{u}\rangle_0} \label{m0}
\end{equation}
together with the result of the parametrization of  $m_{0}$.

Since we have no 't Hooft term, there is no mixing of flavor
sectors, and one can consider the light quark sector independent
of the strange one. The equation for the kaon mass fixes the
strange quark's current mass $m_{s,0}$, whereby a self-consistent
solution of the strange quark gap equation is implied.


\section{Results}

\begin{figure}[th]
\begin{center}
\psfig{figure=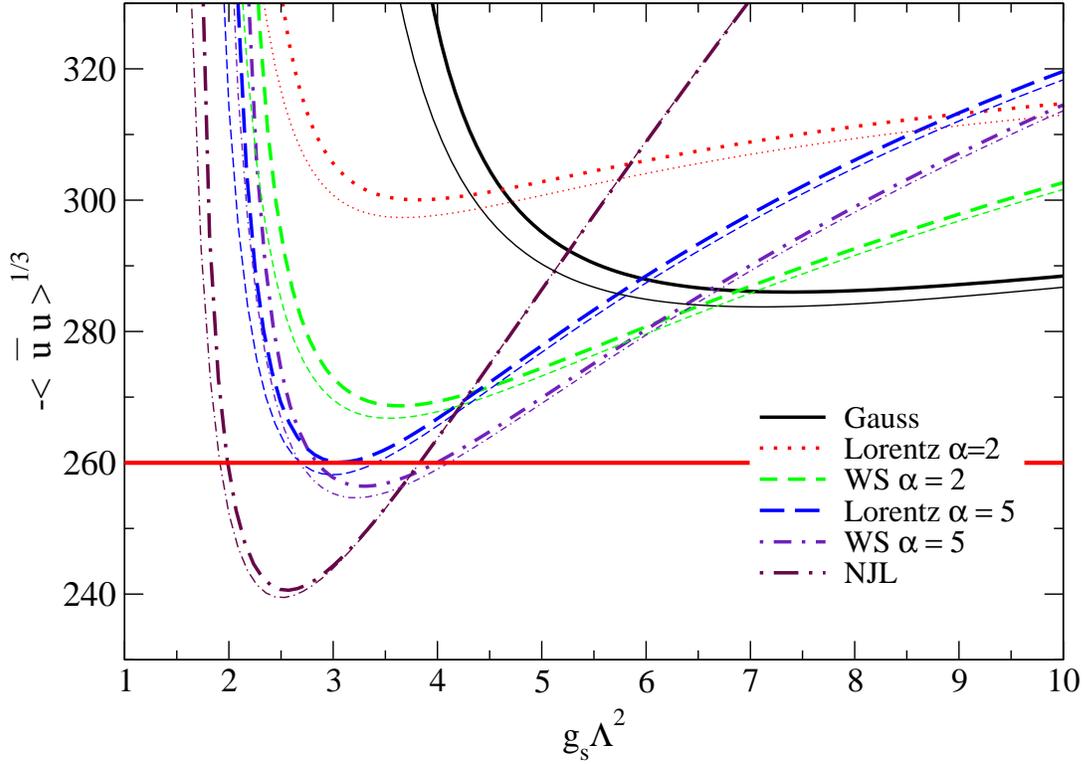,height=0.9\textwidth,angle=-90}
\caption{The dependence of the chiral condensate
$-<\bar{u}u>^{1/3}$ on the choice of the coupling constant in the
chiral limit (thick lines) and for finite current mass of the
light quark $m_{0}\simeq 0.01\Lambda $ (thin lines) for different
formfactor models. The plot demonstrates that not for all cases
the phenomenological constraint for the value of the condensate
from QCD sum rules \protect\cite{Dosch} with an upper limit at
$260$ MeV can be fulfilled.} \label{patalogy}
\end{center}
\end{figure}

In the present parametrization scheme the gap equation plays a
special role. Although the gap is not an observable quantity, we
will use it as an phenomenological input instead of the
condensate, which in some cases does not fulfill the
phenomenological constraints. Moreover, for each formfactor model
there is some minimal value of $G_{S}\Lambda ^{2}$ for which the
condensate has a minimum: for the Gaussian model it is 7.376, for
the Lorentzian model with $\alpha =2$ it is 3.795, and for $\alpha
=10$ it is 2.825. For the NJL model this minimal value is 2.588.
The corresponding values of the condensate are given in Table
\ref{LorsysM}. These values for the finite current masses are
shifted to the left as it is shown in Fig. \ref{patalogy} and for
them the parameter sets are fixed (see Table \ref{LorsysM}). When
the condensate is chosen there are two possible values of
$G_{S}\Lambda ^{2}$ (the lower and higher branches) for which one
can fix the parameters of the model.  We show that the constraint
on the condensate from QCD sum rules \protect\cite{Dosch} with an
upper limit at $260$ MeV can be fulfilled only for values of
$\alpha$ exceeding 3-5 for both Lorentzian and Woods- Saxon
formfactor models  . For the particular choice of $ - <u \bar
u>^{1/3} = 280$ MeV and 260 MeV we fixed the parameters for both
branches of solutions (see Tables \ref{LL260}
-\ref{NJL}).

In order to obtain the parameter sets we choose three values for
the non-observable value of the constituent quark mass
$M(p=0)=330$ MeV, 335 MeV, 367.5 MeV, 380 MeV, and 400 MeV. The
values are taken such that the mass $3M(p=0)$ is larger than the
mass of the nucleon as a bound state of three quarks. The results
of the parametrizations are given in the Tables
\ref{L330}-\ref{L400}.

\begin{figure}[th]
\begin{center}
\psfig{figure=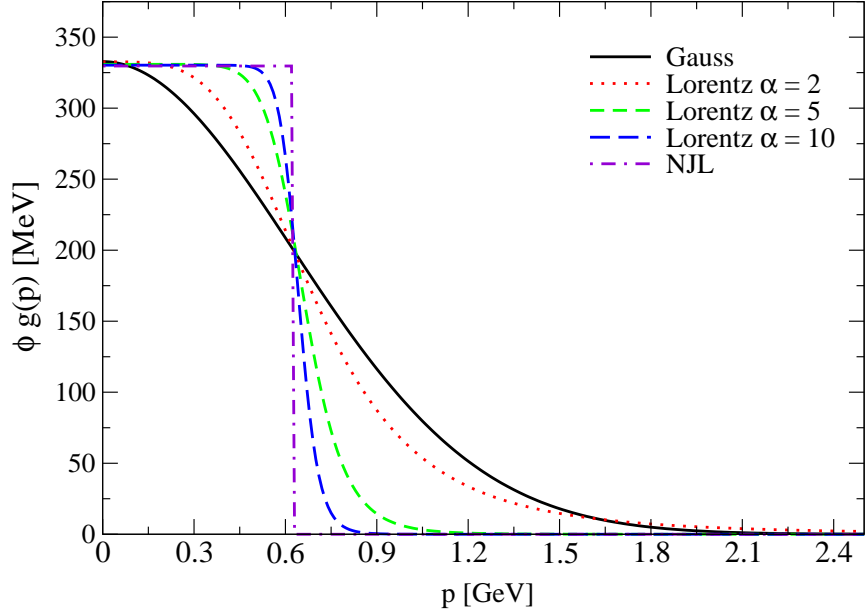,height=0.73\textwidth,angle=-90}
\caption{The gap functions of light quarks for different
formfactor models. The present parameter set is from Table
\ref{L335}. \label{ffcomp}}
\end{center}
\end{figure}

\begin{figure}[th]
\begin{center}
\psfig{figure=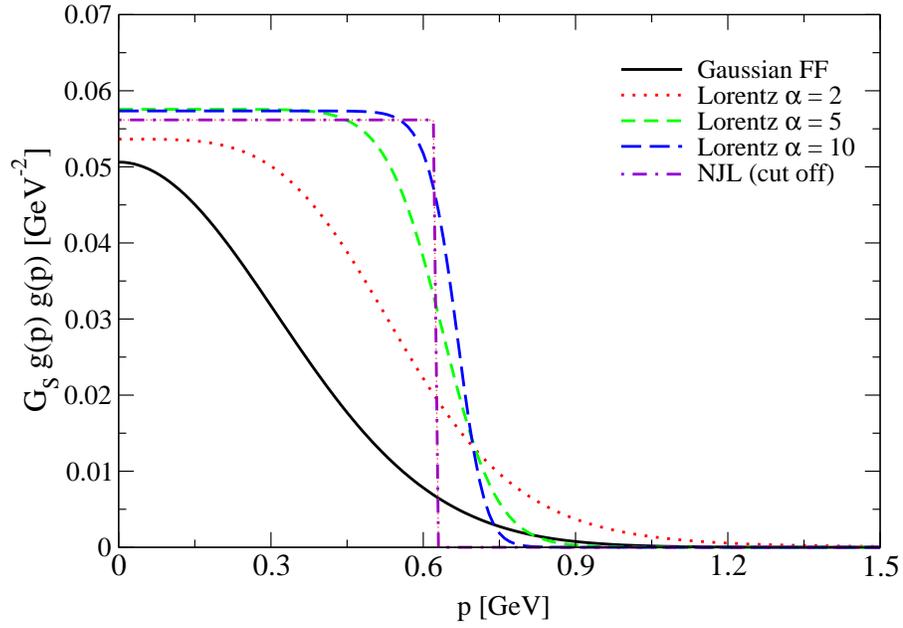,height=0.73\textwidth,angle=-90}
\caption{The diagonal elements of the separable interaction
$G_{S}~g(p)~g(p)$, for different formfactor models. The results
for the parameter set given in Table \ref{L335}. \label{ffpcomp}}
\end{center}
\end{figure}

In Fig. \ref{patalogy} the dependence of the chiral condensate is
shown as a function of $G_S\Lambda^2$ for different formfactors in
the chiral limit band for an appropriate choice of the current
mass $\sim 0.01~\Lambda$. It is shown that the minimal possible
value of the condensate varies from one formfactor model to
another and only in the NJL model the appropriate values of
condensate in the range of QCD sum rule values $230\pm 10$ MeV can
be reached. The Figs. \ref{ffcomp} and \ref{ffpcomp} show the gap
function and the diagonal elements of the separable interaction
for different formfactors in order to demonstrate the systematics
of the changes related to the degree of the softening given by the
parameter $\alpha$ in Lorentzian functions.

\section{Conclusions}

We have presented parametrizations of nonlocal chiral quark models
with instantaneous, separable interactions defined by momentum
dependent formfactors which interpolate between the soft Gaussian
type and the hard cut off (NJL) in tabulated form. The
introduction of a Lorentzian and/or Woods-Saxon-type function with
an additional parameter allowed a systematic investigation of the
NJL model limit, where existing parametrizations could be
recovered.

We have shown that the instantaneous nonlocal models have an
essential problem for the softest formfactors, where it is
impossible to obtain acceptable values for the chiral condensate.
However, for the astrophysical applications this problem could be
considered as of minor importance relative to the insights which a
systematic variation of the interaction model offers for the
better understanding of mechanisms governing the quark matter EoS
on a microscopic level.

We show numerically that the Goldberger-Treiman relation and GMOR
as low-energy theorems hold also for the nonlocal chiral quark
model.

The present approach to nonlocal chiral quark models can be
applied subsequently for systematic studies of constraints on the
EoS of superdense matter coming from the phenomenology of heavy
ion collisions and compact stars.

\subsection*{Acknowledgement}

I thank David Blaschke, Norberto Scoccola and Yuri Kalinovsky for
the initiation of this work and their constructive discussions. I
am grateful to A. Dorokhov, O. Teryaev and V. Yudichev for their
interest in this work and support during my visit at the JINR
Dubna. The research was supported in part by
DFG under grant No. 436 ARM 17/4/05 and by
the DAAD partnership program between the Universities of Rostock
and
Yerevan. 


\bigskip

\begin{table}[bh]
\begin{tabular}{|c|c|c|c|c|c|c|c|c|}
$\alpha $ & - $<u\bar{u}>^{1/3}$ & $m_{0}$ & $m_{0}^{GMOR}$ & $m_{s,0}$ & $%
G_{S}\Lambda ^{2}$ & $\phi _{u,d}$ & $\phi _{s}$ & $\Lambda $ \\
& [MeV] & [MeV] & [MeV] & [MeV] &  & [MeV] & [MeV] & [MeV] \\
\hline\hline G & 329.505 & 2.17720 & 2.17482 & 84.4205 & 3.88079 &
327.823 & 661.678 & 891.044 \\ \hline\hline WS 2 & 287.231 &
3.26230 & 3.28314 & 118.698 & 2.53185 & 326.738 & 567.983 &
681.824 \\ \hline 5 & 264.453 & 4.16354 & 4.20674 & 143.118 &
2.59911 & 325.836 & 502.023 & 680.003 \\ \hline 10 & 253.498 &
4.71893 & 4.77597 & 157.138 & 2.39385 & 325.281 & 467.880 &
653.577 \\ \hline\hline L 2 & 320.280 & 2.36668 &
2.36805 & 86.3960 & 2.57512 & 327.633 & 613.944 & 703.442 \\
\hline 5 & 266.053 & 4.09091 & 4.13132 & 139.122 & 2.49292 &
325.909 & 500.191 & 666.553 \\ \hline 10 & 253.699 & 4.70805 &
4.76458 & 156.362 & 2.35908 & 325.292 & 467.132 & 649.168 \\
\hline\hline $\infty $ & 244.135 & 5.27697 & 5.34681 & 171.210 &
2.17576 & 324.723 & 438.791 & 629.540 \\ \hline
\end{tabular}
\vspace{0.5cm} \caption{Parameter sets for the chiral quark model
with Gaussian,
Woods-Saxon and Lorentzian formfactors for different values of the $\protect%
\alpha $ parameter (including cutoff), when $M(p=0)=330$ MeV.}
\label{L330}
\end{table}

\begin{table}[bh]
\begin{tabular}{|c|c|c|c|c|c|c|c|}
$\alpha $ & - $<u\bar{u}>^{1/3}$ & $m_{0}$ & $m_{s,0}$ &
$G_{S}\Lambda ^{2}$
& $\phi _{u,d}$ & $\phi _{s}$ & $\Lambda $ \\
& [MeV] & [MeV] & [MeV] &  & [MeV] & [MeV] & [MeV] \\ \hline\hline
G & 327.587 & 2.21473 & 85.3586 & 3.90437 & 332.785 & 664.996 &
878.215
\\ \hline\hline
WS 2 & 285.950 & 3.30539 & 119.481 & 2.55036 & 331.695 & 570.557 & 673.959 \\
\hline
5 & 263.581 & 4.20424 & 143.560 & 2.62025 & 330.796 & 504.505 & 673.796 \\
\hline
10 & 252.796 & 4.75777 & 157.365 & 2.41420 & 330.242 & 470.371 & 648.407 \\
\hline\hline L 2 & 319.000 & 2.39420 & 86.9114 & 2.59304 & 332.606
& 617.354 & 695.289 \\ \hline
5 & 265.280 & 4.12606 & 139.418 & 2.51348 & 330.874 & 502.838 & 660.803 \\
\hline
10 & 253.019 & 4.74558 & 156.553 & 2.37921 & 330.254 & 469.656 & 644.106 \\
\hline\hline $\infty $ & 243.515 & 5.31583 & 171.300 & 2.19469 &
329.684 & 441.236 & 625.071 \\ \hline
\end{tabular}
\vspace{0.5cm} \caption{Parameter sets for the chiral quark model
with Gaussian,
Woods-Saxon and Lorentzian formfactors for different values of the $\protect%
\alpha $ parameter (including cutoff), when $M(p=0)=335$ MeV.}
\label{L335}
\end{table}

\begin{table}[bh]
\begin{tabular}{|c|c|c|c|c|c|c|c|}
 $\alpha $ & - $<u\bar{u}>^{1/3}$ & $m_{0}$ & $m_{s,0}$ &
$G_{S}\Lambda ^{2}$
& $\phi _{u,d}$ & $\phi _{s}$ & $\Lambda $ \\
& [MeV] & [MeV] & [MeV] &  & [MeV] & [MeV] & [MeV] \\ \hline\hline
G & 317.279 & 2.43126 & 90.5456 & 4.06161 & 365.069 & 687.275 &
808.486
\\ \hline \hline
WS 2 & 279.325 & 3.54048 & 123.328 & 2.67284 & 363.960 & 588.701 &
631.752
\\ \hline
5 & 259.333 & 4.41036 & 145.100 & 2.75916 & 363.090 & 522.373 & 641.089 \\
\hline
10 & 249.556 & 4.93983 & 157.470 & 2.54737 & 362.560 & 488.449 & 621.664 \\
\hline\hline L 2 & 312.381 & 2.54557 & 89.4547 & 2.71147 & 364.954
& 640.464 & 651.022 \\ \hline
5 & 261.615 & 4.29837 & 140.125 & 2.64838 & 363.202 & 521.702 & 630.566 \\
\hline
10 & 249.906 & 4.92671 & 156.442 & 2.51088 & 362.573 & 487.938 & 617.968 \\
\hline\hline $\infty $ & 240.772 & 5.49540 & 170.417 & 2.31825 &
362.005 & 459.190 & 602.472 \\ \hline
\end{tabular}
\vspace{0.5cm} \caption{Parameter sets for the chiral quark model
with Gaussian,
Woods-Saxon and Lorentzian formfactors for different values of the $\protect%
\alpha $ parameter (including cutoff), when $M(p=0)=367.5$ MeV.}
\label{L367}
\end{table}

\begin{table}[bh]
\begin{tabular}{|c|c|c|c|c|c|c|c|}
$\alpha $ & - $<u\bar{u}>^{1/3}$ & $m_{0}$ & $m_{s,0}$ &
$G_{S}\Lambda ^{2}$
& $\phi _{u,d}$ & $\phi _{s}$ & $\Lambda $ \\
& [MeV] & [MeV] & [MeV] &  & [MeV] & [MeV] & [MeV] \\ \hline\hline
G & 314.115 & 2.50324 & 92.1716 & 4.12364 & 377.497 & 696.149 &
786.678
\\ \hline \hline
WS 2 & 277.402 & 3.61367 & 124.331 & 2.72075 & 376.386 & 596.248 & 618.773 \\
\hline
5 & 258.214 & 4.46967 & 145.205 & 2.81307 & 375.530 & 529.931 & 631.281 \\
\hline
10 & 248.787 & 4.99173 & 157.022 & 2.59883 & 375.008 & 496.137 & 613.860 \\
\hline\hline L 2 & 310.457 & 2.59269 & 90.1287 & 2.75775 & 377.407
& 649.735 & 637.183 \\ \hline
5 & 260.698 & 4.34314 & 139.962 & 2.70064 & 375.657 & 529.610 & 621.523 \\
\hline
10 & 249.185 & 4.97096 & 155.921 & 2.56172 & 375.029 & 495.699 & 610.359 \\
\hline\hline $\infty $ & 240.184 & 5.54297 & 169.559 & 2.36582 &
374.457 & 466.894 & 596.112 \\ \hline
\end{tabular}
\vspace{0.5cm} \caption{Parameter sets for the chiral quark model
with Gaussian,
Woods-Saxon and Lorentzian formfactors for different values of the $\protect%
\alpha $ parameter (including cutoff) , when $M(p=0)=380$ MeV.}
\label{L380}
\end{table}

\begin{table}[bh]
\begin{tabular}{|c|c|c|c|c|c|c|c|}
$\alpha $ & - $<u\bar{u}>^{1/3}$ & $m_{0}$ & $m_{s,0}$ &
$G_{S}\Lambda ^{2}$
& $\phi _{u,d}$ & $\phi _{s}$ & $\Lambda $ \\
& [MeV] & [MeV] & [MeV] &  & [MeV] & [MeV] & [MeV] \\ \hline\hline
G & 309.756 & 2.60719 & 94.4251 & 4.22432 & 397.393 & 710.653 &
756.140
\\ \hline \hline
WS\ 2 & 274.869 & 3.71199 & 125.506 & 2.79810 & 396.288 & 608.881
& 600.815
\\ \hline
5 & 256.870 & 4.53517 & 144.962 & 2.89964 & 395.465 & 542.673 & 617.969 \\
\hline
10 & 247.969 & 5.03928 & 155.906 & 2.68120 & 394.961 & 509.126 & 603.493 \\
\hline\hline L 2 & 307.917 & 2.65479 & 90.9372 & 2.83241 & 397.345
& 664.942 & 617.801 \\ \hline
5 & 259.652 & 4.39512 & 139.341 & 2.78448 & 395.605 & 542.876 & 609.265 \\
\hline
10 & 248.440 & 5.01137 & 154.700 & 2.64310 & 394.989 & 508.802 & 600.271 \\
\hline\hline
 $\infty $ & 239.649 & 5.58218 & 167.771 & 2.44178 &
394.418 & 479.964 & 587.922 \\ \hline
\end{tabular}
\vspace{0.5cm} \caption{Parameter sets for the chiral quark model
with Gaussian, Woods - Saxon and Lorentzian formfactors for
different values of the $\protect\alpha $ parameter (including
cutoff), when $M(p=0)=400$ MeV.} \label{L400}
\end{table}

\begin{table}[bh]
\begin{tabular}{|c|c|c|c|c|c|c|c|c|}
\hline
$\alpha $ & - $<u\bar{u}>^{1/3}$ & $m_{0}$ & $m_{0}^{GMOR}$ & $m_{s,0}$ & $%
G_{S}\Lambda ^{2}$ & $\phi _{u,d}$ & $\phi _{s}$ & $\Lambda $ \\
& [MeV] & [MeV] & [MeV] & [MeV] &  & [MeV] & [MeV] & [MeV] \\
\hline\hline G & 286.005 & 3.2971 & 3.3255 & 103.92 & 6.0252 &
733.24 & 987.97 & 548.85 \\ \hline\hline WS 2 & 268.670 & 3.9696 &
4.0116 & 126.50 & 3.1643 & 489.09 & 674.48 & 546.23
\\ \hline
5 & 256.425 & 4.5602 & 4.6142 & 144.72 & 2.9375 & 404.17 & 548.46
& 612.92
\\ \hline
6 & 254.049 & 4.6875 & 4.7449 & 148.46 & 2.8288 & 393.39 & 530.26
& 615.00
\\ \hline
7 & 252.221 & 4.7859 & 4.8488 & 151.37 & 2.7472 & 386.39 & 517.64
& 615.26
\\ \hline
8 & 250.792 & 4.8765 & 4.9323 & 153.66 & 2.6857 & 381.60 & 508.55
& 614.82
\\ \hline
9 & 249.655 & 4.9418 & 5.0000 & 155.49 & 2.6391 & 378.34 & 501.87
& 614.09
\\ \hline
10 & 248.734 & 4.9959 & 5.0557 & 156.98 & 2.6030 & 376.01 & 496.77
& 613.28
\\ \hline
12 & 247.344 & 5.0798 & 5.1415 & 159.23 & 2.5510 & 373.00 & 489.59
& 611.75
\\ \hline
15 & 245.954 & 5.1659 & 5.2290 & 161.48 & 2.5019 & 370.57 & 482.97
& 609.89
\\ \hline
20 & 244.575 & 5.2543 & 5.3181 & 163.72 & 2.4557 & 368.66 & 476.87
& 607.78
\\ \hline
30 & 243.215 & 5.3318 & 5.4076 & 165.92 & 2.4123 & 367.29 & 471.30
& 605.46
\\ \hline
40 & 242.545 & 5.3772 & 5.4526 & 167.01 & 2.3914 & 366.72 & 468.67
& 604.26
\\ \hline
50 & 242.147 & 5.4046 & 5.4795 & 167.66 & 2.3791 & 366.43 & 467.13
& 603.53
\\ \hline\hline
L 2 & 300.991 & 2.8362 & 2.8531 & 91.830 & 3.2486 & 506.47 &
754.06 & 548.54 \\ \hline 5 & 260.090 & 4.3692 & 4.4219 & 139.68 &
2.7448 & 386.17 & 536.51 & 614.76
\\ \hline
6 & 256.288 & 4.5718 & 4.6218 & 145.31 & 2.6820 & 380.48 & 521.05
& 615.01
\\ \hline
7 & 253.707 & 4.7101 & 4.7641 & 149.24 & 2.6358 & 377.05 & 510.78
& 614.46
\\ \hline
8 & 251.844 & 4.8138 & 4.8706 & 152.13 & 2.6003 & 374.77 & 503.46
& 613.69
\\ \hline
9 & 250.438 & 4.8945 & 4.9532 & 154.33 & 2.5722 & 373.18 & 498.01
& 612.88
\\ \hline
10 & 249.341 & 4.9589 & 5.0190 & 156.07 & 2.5494 & 372.01 & 493.78
& 612.10
\\ \hline
12 & 247.740 & 5.0549 & 5.1166 & 158.62 & 2.5147 & 370.43 & 487.68
& 610.74
\\ \hline
15 & 246.195 & 5.1505 & 5.2135 & 161.10 & 2.4793 & 369.05 & 481.86
& 609.13
\\ \hline
20 & 244.705 & 5.2459 & 5.3096 & 163.51 & 2.4433 & 367.88 & 476.31
& 607.30
\\ \hline
30 & 243.272 & 5.3282 & 5.4039 & 165.83 & 2.4070 & 366.97 & 471.08
& 605.22
\\ \hline
40 & 242.577 & 5.3752 & 5.4505 & 166.96 & 2.3884 & 366.55 & 468.55
& 604.12
\\ \hline
50 & 242.167 & 5.4033 & 5.4782 & 167.62 & 2.3772 & 366.32 & 467.06
& 603.44
\\ \hline\hline
\end{tabular}
\vspace{0.5cm} \caption{Parameter sets for the chiral quark model
with Gaussian,
Woods-Saxon and Lorentzian formfactors for different values of the $\protect%
\alpha $ parameter, when $-<u\bar{u}>^{1/3}$ has its possible
minimal value at chiral limit.} \label{LorsysM}
\end{table}

\begin{table}[bh]
\begin{tabular}{|c|c|c|c|c|c|c||c|c|c|c|c|c|}
&\multicolumn{6}{c||}{The Lower Branch}&\multicolumn{6}{c||}{The
Higher Branch}\\ \hline
$\alpha $ & $m_{0}$ & $m_{s,0}$ & $G_{S}\Lambda ^{2}$ & $\phi _{u,d}$ & $%
\phi _{s}$ & $\Lambda $ & $m_{0}$ & $m_{s,0}$ & $G_{S}\Lambda
^{2}$ & $\phi
_{u,d}$ & $\phi _{s}$ & $\Lambda $ \\
& [MeV] & [MeV] &  & [MeV] & [MeV] & [MeV] & [MeV] & [MeV] &  &
[MeV] & [MeV] & [MeV] \\ \hline \hline 5 & 4.369 & 139.7 & 2.745 &
386.2 & 536.5 & 614.76 & 4.375 & 129.2 & 3.379 & 537.9 & 651.6 &
564.79
\\ \hline 6 & 4.378 & 145.6 & 2.501 & 337.2 & 494.9 & 648.79 &
4.384 & 125.6 & 3.668 & 620.4 & 709.6 & 558.59 \\ \hline 7 & 4.378
& 148.7 & 2.389 & 317.2 & 476.7 & 666.72 & 4.389 & 123.8 & 3.788 &
661.6 & 737.8 & 559.57 \\ \hline 8 & 4.382 & 150.7 & 2.318 & 305.8
& 465.8 & 678.13 & 4.392 & 122.8 & 3.849 & 687.1 & 754.9 & 561.56
\\ \hline 9 & 4.379 & 152.1 & 2.268 & 298.4 & 458.5 & 686.04 &
4.394 & 122.1 & 3.882 & 704.4 & 766.3 & 563.58 \\ \hline 10 &
4.380 & 153.2 & 2.230 & 293.2 & 453.2 & 691.85 & 4.396 & 121.5 &
3.900 & 717.0 & 774.4 & 565.39 \\ \hline 12 & 4.380 & 154.7 &
2.177 & 286.2 & 445.9 & 699.79 & 4.398 & 120.8 & 3.917 & 733.9 &
785.1 & 568.36 \\ \hline 15 & 4.381 & 156.1 & 2.126 & 280.2 &
439.4 & 706.85 & 4.400 & 120.1 & 3.920 & 748.8 & 794.3 & 571.52 \\
\hline 20 & 4.381 & 157.5 & 2.078 & 274.9 & 433.5 & 713.11 & 4.402
& 119.5 & 3.912 & 761.9 & 802.2 & 574.79 \\ \hline 30 & 4.382 &
158.7 & 2.032 & 270.2 & 428.2 & 718.63 & 4.405 & 118.9 & 3.894 &
773.4 & 808.9 & 578.09 \\ \hline 40 & 4.382 & 159.3 & 2.010 &
268.2 & 425.6 & 721.12 & 4.406 & 118.6 & 3.881 & 778.6 & 811.7 &
579.73 \\ \hline 50 & 4.382 & 159.6 & 1.996 & 266.9 & 424.2 &
722.52 & 4.406 & 118.5 & 3.873 & 781.6 & 813.4 & 580.70 \\
\hline
\end{tabular}
\vspace{0.5cm} \caption{Parameter sets for the chiral quark model
with Lorentzian
formfactor for different values of the $\protect\alpha $ parameter, when $-<u%
\bar{u}>^{1/3}=260$ MeV (the both branches).} \label{LL260}
\end{table}

\begin{table}[bh]
\begin{tabular}{|c|c|c|c|c|c|c||c|c|c|c|c|c|}
&\multicolumn{6}{c||}{The Lower Branch}&\multicolumn{6}{c||}{The
Higher Branch}\\ \hline
$\alpha $ & $m_{0}$ & $m_{s,0}$ & $G_{S}\Lambda ^{2}$ & $\phi _{u,d}$ & $%
\phi _{s}$ & $\Lambda $ & $m_{0}$ & $m_{s,0}$ & $G_{S}\Lambda
^{2}$ & $\phi
_{u,d}$ & $\phi _{s}$ & $\Lambda $ \\
& [MeV] & [MeV] &  & [MeV] & [MeV] & [MeV] & [MeV] & [MeV] &  &
[MeV] & [MeV] & [MeV] \\ \hline\hline 5 & 3.522 & 131.0 & 2.278 &
273.2 & 477.9 & 758.35 & 3.520 & 97.42 & 5.256 & 1008 & 1076 &
546.27 \\ \hline 6 & 3.523 & 133.7 & 2.198 & 261.8 & 464.5 &
779.62 & 3.522 & 96.57 & 5.294 & 1041 & 1096 & 558.38 \\ \hline 7
& 3.523 & 135.4 & 2.143 & 255.1 & 456.2 & 792.89 & 3.525 & 96.03 &
5.281 & 1058 & 1105 & 567.40 \\ \hline 8 & 3.523 & 136.7 & 2.103 &
250.7 & 450.6 & 801.86 & 3.524 & 95.64 & 5.252 & 1067 & 1110 &
574.24 \\ \hline 9 & 3.526 & 137.6 & 2.072 & 247.6 & 446.5 &
808.28 & 3.526 & 95.35 & 5.219 & 1073 & 1112 & 579.55 \\ \hline 10
& 3.525 & 138.3 & 2.047 & 245.3 & 443.4 & 813.06 & 3.532 & 95.13 &
5.187 & 1077 & 1113 & 583.78 \\ \hline 12 & 3.524 & 139.4 & 2.011
& 242.1 & 439.0 & 819.66 & 3.527 & 94.80 & 5.129 & 1081 & 1113 &
590.05 \\ \hline 15 & 3.527 & 140.3 & 1.975 & 239.1 & 434.8 &
825.61 & 3.528 & 94.48 & 5.062 & 1084 & 1112 & 596.21 \\ \hline 20
& 3.524 & 141.3 & 1.939 & 236.5 & 431.0 & 830.90 & 3.530 & 94.17 &
4.987 & 1086 & 1110 & 602.20 \\ \hline 30 & 3.525 & 142.1 & 1.903
& 234.0 & 427.3 & 835.56 & 3.533 & 93.86 & 4.904 & 1087 & 1107 &
608.00 \\ \hline 40 & 3.526 & 142.5 & 1.885 & 232.9 & 425.5 &
837.65 & 3.535 & 93.69 & 4.860 & 1087 & 1106 & 610.81 \\ \hline 50
& 3.526 & 142.8 & 1.874 & 232.2 & 424.5 & 838.83 & 3.537 & 93.60 &
4.833 & 1087 & 1105 & 612.47 \\ \hline
\end{tabular}
\vspace{0.5cm} \caption{Parameter sets for the chiral quark model
with Lorentzian
formfactor for different values of the $\protect\alpha $ parameter, when $-<u%
\bar{u}>^{1/3}=280$ MeV (for both branch).} \label{LL280}
\end{table}

\begin{table}[bh]
\begin{tabular}{|c|c|c|c|c|c|c||c|c|c|c|c|c|}
&\multicolumn{6}{c||}{The Lower Branch}&\multicolumn{6}{c||}{The
Higher Branch}\\ \hline
$\alpha $ & $m_{0}$ & $m_{s,0}$ & $G_{S}\Lambda ^{2}$ & $\phi _{u,d}$ & $%
\phi _{s}$ & $\Lambda $ & $m_{0}$ & $m_{s,0}$ & $G_{S}\Lambda
^{2}$ & $\phi
_{u,d}$ & $\phi _{s}$ & $\Lambda $ \\
& [MeV] & [MeV] &  & [MeV] & [MeV] & [MeV] & [MeV] & [MeV] &  &
[MeV] & [MeV] & [MeV] \\ \hline \hline 5 & 4.380 & 145.0 & 2.732 &
356.7 & 518.6
& 646.62 & 4.386 & 125.4 & 4.084 & 672.1 & 762.0 & 554.94 \\
\hline 6 & 4.378 & 147.9 & 2.560 & 330.5 & 493.1 & 666.14 & 4.389
& 123.9 & 4.108 & 698.3 & 775.6 & 558.34 \\ \hline 7 & 4.378 &
150.0 & 2.446 & 315.1 & 477.4
& 678.52 & 4.391 & 122.9 & 4.097 & 714.3 & 783.0 & 561.32 \\
\hline 8 & 4.374 & 151.6 & 2.366 & 305.0 & 467.0 & 686.93 & 4.393
& 122.2 & 4.078 & 725.2 & 787.9 & 563.77 \\ \hline 9 & 4.379 &
152.7 & 2.307 & 298.0 & 459.5 & 692.98 & 4.395 & 121.6 & 4.058 &
733.3 & 791.4 & 565.79 \\ \hline 10 & 4.380 & 153.7 & 2.263 &
292.9 & 454.0 & 697.49 & 4.396 & 121.2 & 4.040 & 739.6 & 794.1 &
567.46 \\ \hline 12 & 4.380 & 155.0 & 2.199 & 286.1 & 446.5 &
703.75 & 4.398 & 120.6 & 4.010 & 748.9 & 798.2 & 570.06 \\ \hline
15 & 4.380 & 156.3 & 2.141 & 280.1 & 439.8 & 709.43 & 4.400 &
120.0 & 3.979 & 757.9 & 802.3 & 572.76 \\ \hline 20 & 4.381 &
157.6 & 2.086 & 274.8 & 433.7 & 714.60 & 4.402 & 119.4 & 3.944 &
766.8 & 806.4 & 575.56 \\ \hline 30 & 4.382 & 158.7 & 2.035 &
270.1 & 428.2 & 719.30 & 4.405 & 118.9 & 3.907 & 775.5 & 810.6 &
578.46 \\ \hline 40 & 4.382 & 159.3 & 2.011 & 268.1 & 425.7 &
721.50 & 4.406 & 118.6 & 3.889 & 779.8 & 812.7 & 579.95 \\ \hline
50 & 4.382 & 159.7 & 1.997 & 266.8 & 424.2 & 722.77 & 4.400 &
118.7 & 3.877 & 782.4 & 814.0 & 580.85 \\ \hline
\end{tabular}
\vspace{0.5cm} \caption{Parameter sets for the chiral quark model
with Woods-Saxon
formfactor for different values of the $\protect\alpha $ parameter, when $-<u%
\bar{u}>^{1/3}=260$ MeV (for both branches).} \label{WS260}
\end{table}

\begin{table}[bh]
\begin{tabular}{|c|c|c|c|c|c|c||c|c|c|c|c|c|}
&\multicolumn{6}{c||}{The Lower Branch}&\multicolumn{6}{c||}{The
Higher Branch}\\ \hline
$\alpha $ & $m_{0}$ & $m_{s,0}$ & $G_{S}\Lambda ^{2}$ & $\phi _{u,d}$ & $%
\phi _{s}$ & $\Lambda $ & $m_{0}$ & $m_{s,0}$ & $G_{S}\Lambda
^{2}$ & $\phi
_{u,d}$ & $\phi _{s}$ & $\Lambda $ \\
& [MeV] & [MeV] &  & [MeV] & [MeV] & [MeV] & [MeV] & [MeV] &  &
[MeV] & [MeV] & [MeV] \\ \hline\hline 2 & 3.516 & 123.0 & 2.658 &
360.0 & 586.4 & 636.21 & 3.519 & 98.88 & 6.018 & 1223 & 1324 &
464.2 \\ \hline 5 & 3.521 & 132.9 & 2.378 & 272.3 & 481.4 & 780.90
& 3.528 & 96.37 & 6.024 & 1155 & 1209 & 554.4 \\ \hline 6 & 3.522
& 134.7 & 2.280 & 262.4 & 468.2 & 795.37 & 3.521 & 95.93 & 5.824 &
1138 & 1185 & 565.5 \\ \hline 7 & 3.523 & 136.1 & 2.208 & 255.8 &
459.1 & 804.73 & 3.524 & 95.60 & 5.664 & 1127 & 1169 & 573.4 \\
\hline 8 & 3.526 & 137.1 & 2.154 & 251.3 & 452.8 & 811.16 & 3.525
& 95.34 & 5.538 & 1118 & 1156 & 579.2 \\ \hline 9 & 3.524 & 137.9
& 2.112 & 248.0 & 448.2 & 815.77 & 3.526 & 95.12 & 5.440 & 1112 &
1148 & 583.7 \\ \hline 10 & 3.524 & 138.6 & 2.080 & 245.6 & 444.7
& 819.24 & 3.527 & 94.95 & 5.362 & 1108 & 1141 & 587.3 \\ \hline
12 & 3.524 & 139.5 & 2.033 & 242.2 & 439.8 & 824.08 & 3.535 &
94.68 & 5.247 & 1102 & 1132 & 592.7 \\ \hline 15 & 3.524 & 140.4 &
1.989 & 239.2 & 435.3 & 828.51 & 3.528 & 94.41 & 5.135 & 1097 &
1123 & 598.0 \\ \hline 20 & 3.524 & 141.3 & 1.947 & 236.5 & 431.2
& 832.57 & 3.530 & 94.13 & 5.026 & 1093 & 1116 & 603.3 \\ \hline
30 & 3.525 & 142.2 & 1.906 & 234.0 & 427.4 & 836.31 & 3.533 &
93.85 & 4.920 & 1089 & 1110 & 608.5 \\ \hline 40 & 3.526 & 142.6 &
1.887 & 232.9 & 425.6 & 838.08 & 3.535 & 93.68 & 4.869 & 1088 &
1107 & 611.1 \\ \hline 50 & 3.526 & 142.8 & 1.875 & 232.2 & 424.5
& 839.12 & 3.537 & 93.59 & 4.839 & 1088 & 1106 & 612.7 \\ \hline
\end{tabular}
\vspace{0.5cm} \caption{Parameter sets for the chiral quark model
with Woods-Saxon
formfactor for different values of the $\protect\alpha $ parameter, when $-<u%
\bar{u}>^{1/3}=280$ MeV (for both branches).} \label{WS280}
\end{table}
\begin{table}[bh]
\begin{tabular}{|c|c|c|c|c|c|c||c|c|c|c|c|c|}
&\multicolumn{6}{c||}{The Lower Branch}&\multicolumn{6}{c||}{The
Higher Branch}\\ \hline
 $<u\bar{u}>^{1/3}$ & $m_{0}$ & $m_{s,0}$ & $G_{S}\Lambda ^{2}$ &
$\phi _{u,d}$ & $\phi _{s}$ & $\Lambda $ & $m_{0}$ & $m_{s,0}$ &
$G_{S}\Lambda ^{2}
$ & $\phi _{u,d}$ & $\phi _{s}$ & $\Lambda $ \\
\lbrack MeV] & [MeV] & [MeV] &  & [MeV] & [MeV] & [MeV] & [MeV] &
[MeV] &  & [MeV] & [MeV] & [MeV] \\ \hline\hline -280.0 & 3.5267 &
143.7 & 1.8312 & 229.7 & 420.6 & 842.995 & 3.5212 & 93.24 & 4.7187
& 1085 & 1200 & 618.877 \\ \hline -260.0 & 4.3832 & 160.9 & 1.9429
& 262.3 & 418.7 & 727.552 & 4.4026 & 117.8 & 3.8296 & 792.3 &
819.0 & 584.552 \\ \hline -250.8 & 4.8723 & 168.0 & 2.0424 & 289.5
& 424.6 & 672.837 & 4.9032 & 133.1 & 3.3873 & 657.7 & 695.2 &
572.204 \\ \hline -247.5 & 5.0661 & 169.9 & 2.0970 & 304.0 & 429.7
& 652.141 & 5.0903 & 139.7 & 3.2119 & 606.4 & 649.5 & 569.133 \\
\hline -242.4 & 5.3922 & 171.3 & 2.2346 & 340.1 & 446.7 & 616.631
& 5.4081 & 152.2 & 2.8899 & 515.3 & 571.8 & 568.224 \\ \hline
-240.8 & 5.5052 & 170.4 & 2.3161 & 361.4 & 458.9 & 602.778 &
5.5093 & 157.5 & 2.7509 & 477.1 & 541.0 & 570.738 \\ \hline
\end{tabular}
\vspace{0.5cm} \caption{Parameter sets for the chiral quark model
with NJL formfactor for different values of the condensate
$-<u\bar{u}>^{1/3}$ (for both branches)} \label{NJL}
\end{table}

\end{document}